**Charting the Course for Elementary Particle Physics**

**AAAS Panel, San Francisco, CA**

**Prof. Burton Richter**

**Stanford University**

**February 16, 2007**

## INTRODUCTION

"It was the best of times; it was the worst of times" is the way Dickens begins the Tale of Two Cities.  The line is appropriate to our time in particle physics.  It is the best of times because we are in the midst of a revolution in understanding, the third to occur during my career.  It is the worst of times because accelerator facilities are shutting down before new ones are opening, restricting the opportunity for experiments, and because of great uncertainty about future funding.  My task today is to give you a view of the most important opportunities for our field under a scenario that is constrained by a tight budget. It is a time when we cannot afford the merely good, but must give first priority to the really important.

The defining theme of particle physics is to learn what the universe is made of and how it all works. This definition spans the full range of size from the largest things to the smallest things.  This particle physics revolution has its origins in experiments that look at both.

The first of my three revolutions occurred in 1950s when I was a student. It was marked by the discovery of more and more meson and baryon resonances that we then unhappily thought to be elementary and whose names filled the Particle Data Book. It also was the time when parity was found not to be conserved, my first experience with the overthrow of a faith based theory unsupported by experiment. The tools of the time were mainly the bubble chambers using secondary beams from fixed target accelerators.

The second revolution began in the late 60s, and by the mid-70s culminated in the establishment of the Standard Model. Everyone knew even then that the Standard

Work supported in part by the US Department of Energy contract DE-AC02-76SF00515.

Model was only valid in a limited energy range. From the middle of the 1970s to the middle of the 1990s we tried to find what was beyond the Standard Model. The two big questions were where did it break down and what would supersede it. The tools included big electronic tracking chambers and colliding beam machines.

Over the last 15 years we have found that we knew much less about what the universe is made of and how it works than we thought we did. Neutrinos could change from one type to another. Visible matter only made up 5% of the energy density of our universe; dark matter made up 25%; and something called dark energy made up the remaining 70%. The tools now included telescopes and satellites and underground facilities in addition to the big accelerators.

The next 10 to 15 years will answer many questions and raise new ones. We may find what is beyond the Standard Model, what at least some of the dark matter is made of, and what is driving the apparently accelerating expansion of the universe. It may even see an experimental test of higher dimensions and of string theory.

Regrettably, the experiments are bigger and more expensive than the last round and so finance will limit the pace of discovery. My personal priority list includes some that are sure to get done, and some that we ought to get done. I will discus:

LHC and its possible upgrades, including the LHC's impact on the ILC;

ILC, including parameters, schedule, budget and next steps;

Accelerator R&D and its importance for the future;

Dark Matter and Dark Energy searches from the ground and space including cosmic rays of all kinds;

Neutrinos from reactors, accelerators, as well as double beta decay.

## LHC

The late 1980s saw the beginning of the Superconducting Super Collider (SSC), a 40 TeV proton-proton colliding beam facility. The year 1993 saw its end. The SSC was replaced with a more modest machine, the 14 TeV Large Hadron Collider (LHC) at CERN. With its lower energy and its use of the infrastructure of the 27 km circumference LEP electron-positron collider, the CERN facility could be built at much less cost than the SSC. The prospect was attractive and contributions from outside the CERN member states flowed in.  While CERN is not a world lab, the LHC is a world machine.

**Table 1: LHC schedule**

| | |
|---|---|
| 2007 (4th Quarter) | – Collisions at 450×450 GeV |
| 2008  (1st Quarter) | – Shut down to ramp up energy and finish detectors |
| 2008 (2nd Quarter) | – Begin full operations and commission & calibrate detectors |
| 2008 (2nd Quarter) | – Request to council for funds to begin work on 10×**L** upgrade |

The LHC commissioning schedule has already been discussed at this meeting (Table 1).  Colliding beams at or near top energy are expected in the summer of 2008 and physics results should begin to appear in 2009. Compared to today's highest energy facility, the FNAL Tevatron, the LHC has 100 times the luminosity and 7 times the energy. Thus, the LHC needs only a few percent of its design luminosity to see in its first experimental year a low-mass Higgs, if one is there as expected. Other things, like super symmetry, will take longer to check. What we all hope for is one of those illuminating surprises that change the course of science. The physics that will come out in the first few years will have a profound effect on the prospects of the Linear Collider.

The LHC detectors are awesome (Figure 1).  Peter Jenni, the Atlas spokesman, worked with me on the Mark I detector at SPEAR.  It is interesting to see how far detectors and Peter have come since his post-doc days (Figure 2).

**Figure 1: ATLAS Detector**

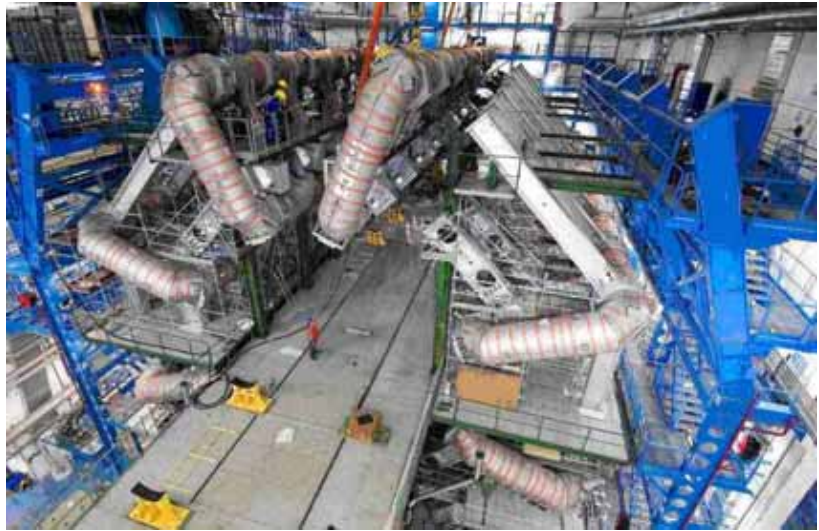

(Photo courtesy of CERN)

**Figure 2: LHC ATLAS Detector and the SPEAR Mark I Detector to scale**

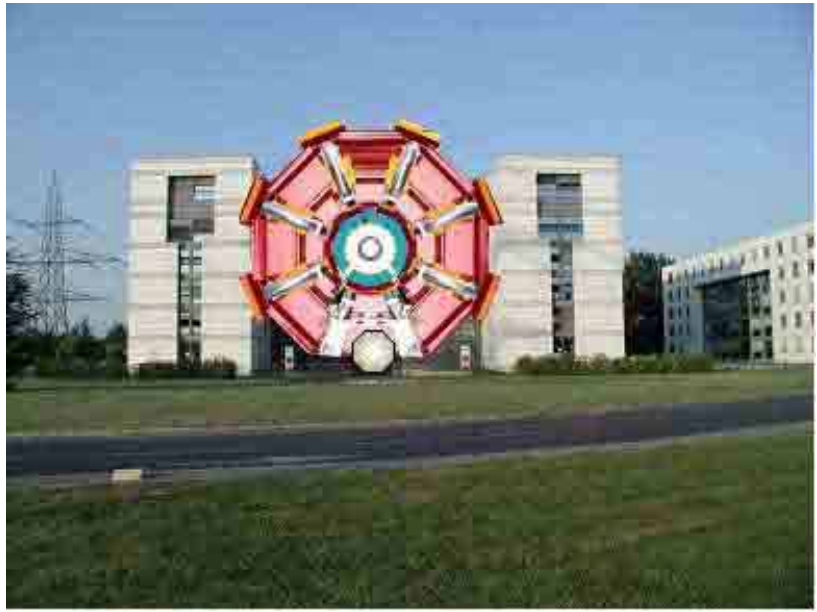

An upgrade is already being planned for the LHC aimed at an increase in its luminosity by a factor of 10 which will roughly double the LHC's mass reach. The upgrade will require rebuilding parts of the LHC and its injector complex.  In addition, major parts of the detectors' tracking systems will have to be replaced to handle the increased radiation form the beam-beam interaction and the increased event rate which will go from 25 to 250 events per beam crossing at full luminosity. The CERN Council will be asked to increase the budget beginning in 2008 for this program. If all goes well, the shutdown for the upgrade will be about 5 years after the start of full operations.

**ILC**

Linear colliders were born at a conference in 1978. The International Committee for Future Accelerators (ICFA) held a workshop at Fermilab on the Limits in Energy of Particle Accelerators. At that workshop, Alexander Skrinski of Novosibirsk, Maury Tigner of Cornell, and I discovered that we had all been thinking about a new kind of electron-positron colliding-beam device, a linear collider. The first of these, a kind of folded linear collider, was built at SLAC. It was not until this machine began to produce real physics with polarized beams at reasonable luminosity that the world began to take linear colliders seriously.

Today, a global design effort aimed at the realization of a machine is underway. It is the end product of a world-wide R&D program that Hirotaka Sugawara, then Director General of KEK, Bjorn Wiik, then the Director of DESY, and I began. We did this because of the lesson of the SSC. If a true world accelerator project was to be built, potential partners should be involved from the very beginning.

On Feb. 8, 2007 the Global Design Effort (GDE) under the leadership of Barry Barish unveiled its first Reference Design Report (RDR) for a superconducting ½ TeV linear collider, upgradeable in the future to 1 TeV (Figure 3). The design luminosity is $2\times10^{34}$ and it accommodates 2 detectors in a single large hall. A future upgrade to 1 TeV requires the addition of 10 km of tunnel and superconducting accelerator at each end.

**Figure 3: Baseline configuration of the ILC**

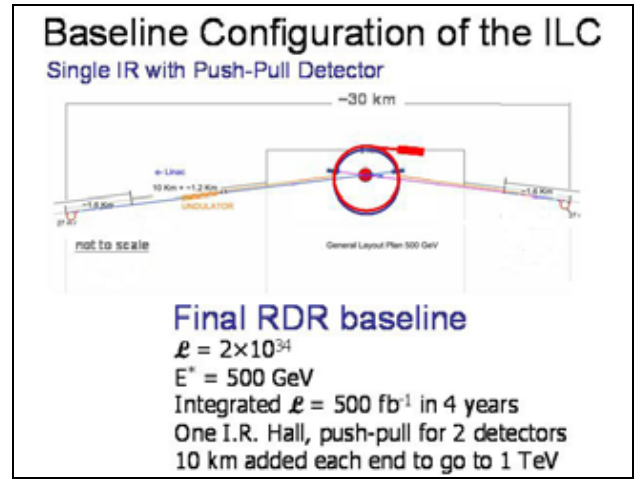

The schedule (Figure 4) believed possible is to begin the detailed engineering design in 2008, complete it in 2010, start construction in 2012, and begin operations in 2019. This earliest possible turn on time will be 10 years after the LHC begins to generate its physics output.

**Figure 4: ILC overall timeline**

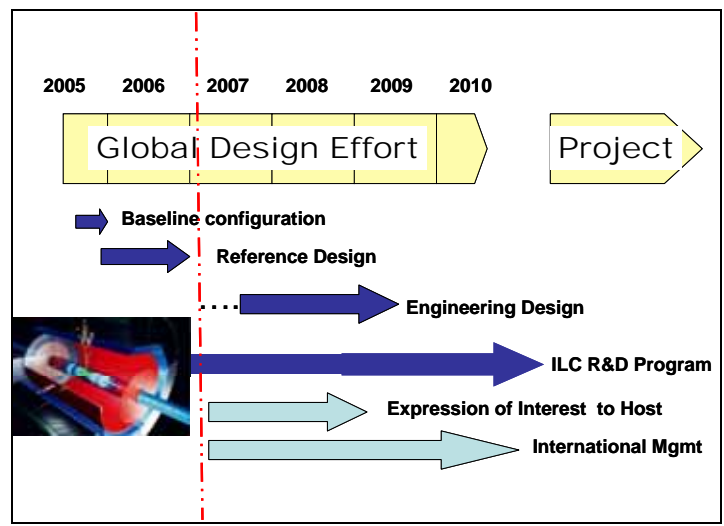

The cost (Table 2) of the ½ TeV version is estimated to be $6.7 Billion in procurements for the technical plus conventional facilities plus 13,000 person-years of additional labor (about $1.3 Billion) all in 2007 dollars. The additional labor is for the people who do the engineering, testing, etc. In the U.S. these are normally lab people and are included in the budget for a project. In Europe and Asia they are typically not included. An international technical experts group will review the estimate in a few months. To get costs in the style the U.S. government typically uses, a rough rule of thumb is to take the procurement estimate, add inflation, and double it. Of course this is an international project and the costs will be shared. This budget and people estimate does not include detectors or any facilities required to support the experimental program (offices, labs, computing, cafeteria, etc).

**Table 2: Costs from the ILC Design Report ("Value Method")**

| | |
|---|---|
| Accelerator systems | Not Included: |
| $6.6 (Billion '07) | ✎ Escalation |
| Personnel | ✎ Contingency |
| | ✎ Preconstruction R&D |
| 13,000 person years | ✎ Detectors |
| | ✎ Support facilities for physics |
| | ✎ Commissioning |

The GDE group has pulled off a near miracle to get as far as they have as a purely volunteer organization. They have no money directly and no control over the expenditures on R&D of the participating institutions. The design is a testament to inspired leadership and to good will on the part of the participants. However, this kind of organization cannot carry out the next stage, the detailed engineering design that will produce what we would call the Technical Design Report. To do that requires an organization that controls the funds, sets priorities for the R&D, oversees the work, and is accountable to the agencies that are paying the bills. The International funding agencies have to get serious and create a real organization. Beyond this stage is the creation of the organization that will build and operate the facility. That is for another day.

I believe that if the LHC works well and finds nothing by 2012, the linear collider will not be funded. It is much more likely that the LHC will produce a great deal of physics during the 10 year period between the LHC first physics output and the ILC's start up. There may be great pressure to turn it on at its full 1 TeV energy. I have no idea what that would do to the budget.

## ACCELERATOR R&D

If we are not to see the end of the era of accelerator based High Energy Physics in the next few decades, we will have to invest more in accelerator R&D.  The big proton machines are already approaching a refrigeration limit in handling synchrotron radiation at liquid helium temperatures.  The radiated power goes as the square of the energy times the square of the magnetic field.  A factor of 2 to 3 in mass reach will require a factor of about 10 in refrigeration even at a constant magnetic field and much more if the field goes up as well.  It is time to get to work on high temperature superconducting magnets if proton machines are to keep going.

There seems to be more room for energy increases at electron machines.  The CLIC group at CERN (Figure 5) is working on the technology for a 3 TeV linear collider. They have always used room temperature accelerating structures, and have recently changed RF frequencies to X-band (12 GHz), the same frequency of the KEK-SLAC ILC room temperature option. They hope for 100 MV/m accelerating gradient, not far above the 60 MV/m achieved a few years ago.

**Figure 5: CLIC schematic (as of Dec. 2006)**

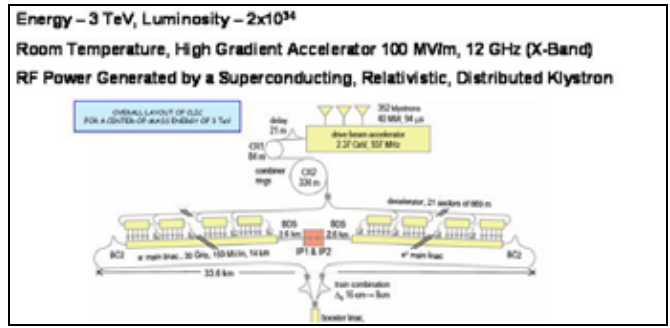

There are exciting things going on in plasma wake-field and laser acceleration. In the Feb. 15 issue of "Nature" magazine, the results of a plasma wake-field acceleration experiment have been published (Figure 6). They demonstrate 50 GV/m acceleration, more than 1000 times the gradient proposed for the ILC. While the efficiency and spectrum demonstrated in this experiment are poor, this is a beginning for something with real potential. High gradients have also been demonstrated in laser acceleration where 2 stages of acceleration have been demonstrated.

This kind of work is among the truly important.

**Figure 6.**

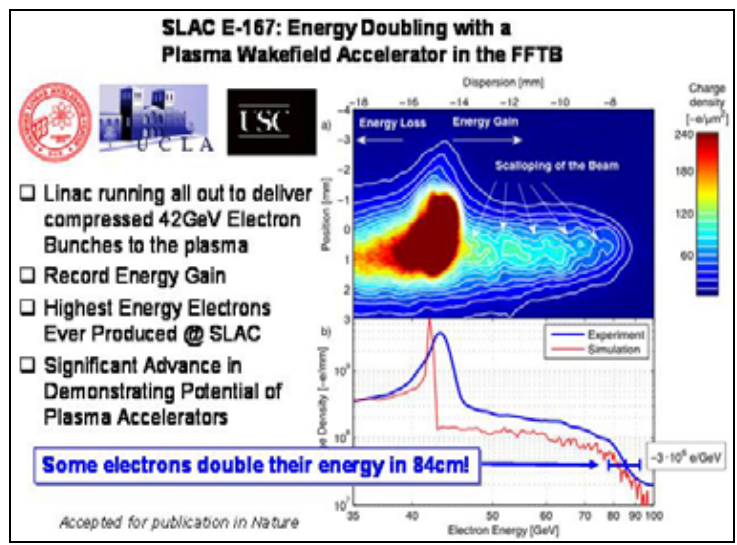

## ASTRO-PARTICLE PHYSICS

While the big accelerators have been trying to find what is beyond the Standard Model, results from non-accelerator experiments have been transforming our view of the universe's content and workings. Davis' experiments showing the solar neutrino deficit and Koshiba's development of the giant Kamiokande detectors led to the revelation of neutrino oscillations. Mather and Smoot with the COBE experiment followed up by WMAP, and Perlmutter, Reiss, and Schmidt with their supernova studies have transformed our notions of the content of the universe (Table 3). It is not so easy to remain anthropically focused when you find that the stuff that we are composed of only makes up roughly 5% of the energy density of our universe.

**Table 3.**

| Dark Matter (25%), Dark Energy (70%), and Us (5%) |
| :--- |
| The 5%:  Has occupied almost all of our attention. |
| The 25%:  Zwicky in the 1930s (velocities in galaxy clusters)  Rotation curves of stars in Galaxy WIMPS. |
| The 70%:  A surprise from the SN1A search. |

Dark matter has been on the science table since the 1930s when Zwicky used the idea to explain the relative velocity of galaxies in clusters. Dark energy came as a complete surprise with the first results on the distance-luminosity ratio of type SN1A supernova. Today's universe is quite different from that of only 10 years ago.

The most dramatic demonstration of the existence of dark matter (Figure 7) comes from the Bullet cluster. The figure is a composite of data from the Hubble space telescope and the Chandra x-ray satellite. The red shows the x-rays measured by Chandra. The blue shows the invisible mass as determined from gravitational lensing. Two giant clusters of galaxies collided long ago. The weakly interacting dark matter in each cluster passed through each other with little if any interactions. The ordinary matter interacted, and that interaction generated a drag that slowed and heated it, generating the x-rays.

**Figure 7: Composite picture from Hubble, Chandra, and lensing data**

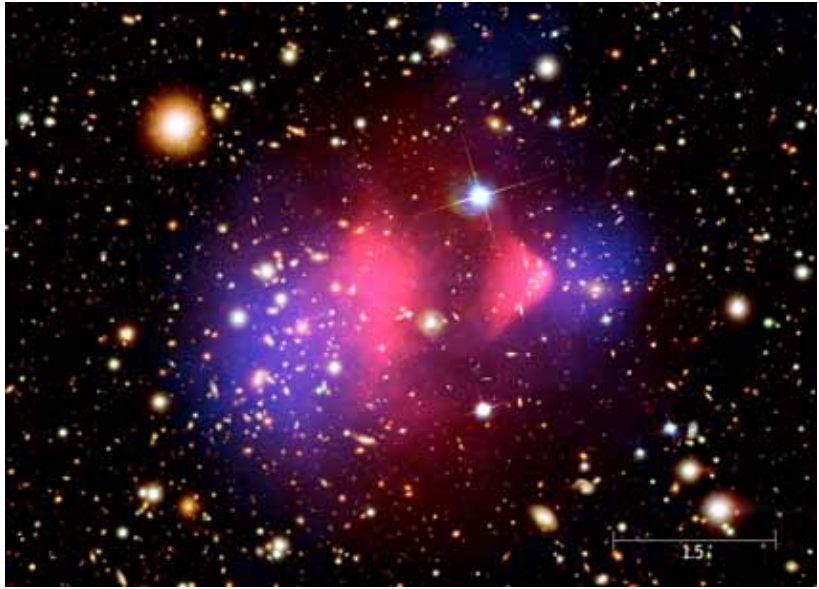

(Credit: X-ray: NASA/CXC/CfA/M.Markevitch et al.;

Optical: NASA/STScI; Magellan/U.Arizona/D.Clowe et al.;

Lensing Map: NASA/STScI; ESO WFI; Magellan/U.Arizona/D.Clowe et al.)

What we know about dark matter from the astro-particle results strongly constrains the properties of candidate super symmetry models. The mass and properties couplings of such things cannot be such as to generate too much dark matter.

The supernova surveys continue to generate more data. The most recent come from Hubble data and extend to higher "z" (Figure 8). The goal is to determine the equation of state parameter and its time evolution. That takes much more data than there is so far.

Figure 8.

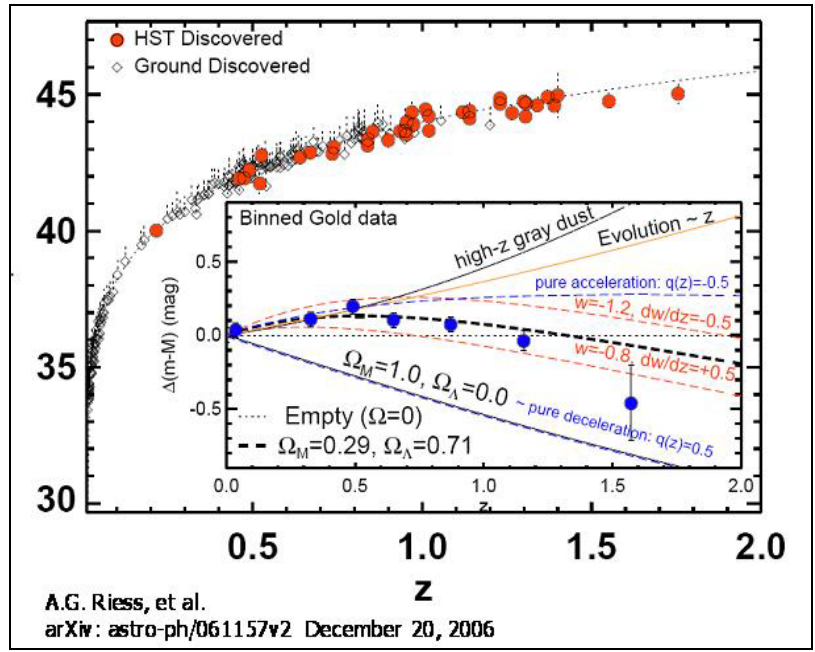

A.G. Riess, et al.
arXiv: astro-ph/061157v2  December 20, 2006

New experiments are starting or are soon to start.  CDMS is hunting for dark matter underground. The GLAST and PLANCK satellites are soon to launch.  VERITAS and Auger are studying ultra-high energy cosmic rays.  The SLOAN Digital Sky survey continues to collect data.  From the astro-particle perspective, however, there are two more very important programs that are yet to be funded.  The first is the LSST, a ground-based optical telescope to be jointly funded by the DOE and NSF.  The DOE is ready to go, but the NSF has competing projects from the astronomy community that need to be prioritized.

The second project is the Joint Dark Energy Mission (JDEM), a DOE-NASA project.  Here, too the problems come from outside the DOE.  NASA's budget is under great pressure from the SPACE Station and the Moon-Mars program.  There is an Academy panel working now to set priorities for the NASA science program.

It is going to be an exciting decade.

## THE FOUR NEUTRINO QUESTIONS

The first of my four neutrino questions is on the existence of a fourth generation, sterile neutrino. The Los Alamos (LSND) experiment gave a positive result, but that result has never been confirmed, gave significant theory problems, and generated some doubts about backgrounds. The MiniBooNE experiment at Fermilab is designed to give a definitive yes or no to a signal in the range of the LSND result. However, it is taking longer to come out with a result than was originally forecast.

The experimenters are doing a blind analysis and will "open the box" sometime soon. I never did like blind analyses. Today's blind experiments are done to prevent unconscious biases from biasing the analysis toward some unconsciously desired result. It does that, but at the cost of preventing the experimenters from seeing problems in the data and certainly prevents them from following clues to new and unexpected results. I always wanted to check on the signal before worrying about the background, and you cannot do that in a blind analysis.

Since we do not have an answer from MiniBooNE as yet, we should consider the possibility that they will not have a conclusive result. In that case another experiment will be needed and this one should be done at the Oak Ridge Spallation Neutron Source. The energy of the machine is below K meson threshold, reducing background. The beam pulse is short, eliminating neutron diffusion as a background source. The beam power is higher than any other machine available today. I very much hope that we do not need another shot at the answer, but if we do, the best source is available.

The next question on my list is the size of $\theta_{13}$. It is interesting in its own right, but if it is too small CP violation in the neutrino sector will be unmeasurable. The present limit of $\sin^2(2\theta_{13})$ of less than 0.1 comes from the CHOOZ reactor experiment. New experiments with reactors (CHOOZ-2 and Daya Bay) and accelerators (JPARC) should be sensitive to $\theta_{13}$ down to an upper bound of $\sin^2(2\theta_{13})$ of 0.01. If it is not larger than this, I doubt that CP violation can be measured. It would be wise to get the new results on $\theta_{13}$ before embarking on other large and costly accelerator based neutrino experiments.

The last two questions, the Majorana or Dirac nature of the neutrino and the normal or inverted mass hierarchy, are loosely linked. If the mass scheme is inverted, the electron neutrino mass must be greater than 0.05 eV, and the next generation of double beta decay experiments could see an effect if the neutrino was a Majorana particle. If the mass hierarchy is normal, the electron neutrino mass can be much smaller and I doubt that neutrinoless double beta decay can ever be observed.

**A FINAL WORD**

The way forward as laid out here is based on the signposts planted by the experiments that have gone before. The next 5 years will plant new signposts from accelerator and non-accelerator experiments. Those may point in different directions and a better way forward may be created. Don't preserve all of the options beyond their "sell by" date.